\documentclass[twoside,12pt]{article}
\usepackage{graphicx}
\begin{document}
\centerline{\Large \bf Bit-String Models for Parasex}
\bigskip
S. Moss de Oliveira$^1$, P.M.C. de Oliveira$^1$ and D. Stauffer$^2$.

\bigskip
Laboratoire PMMH, Ecole Sup\'erieure de Physique et Chimie
Industrielle, 10 rue Vauquelin, F-75231 Paris, Euroland

\medskip

\noindent
$^1$ Visiting from Instituto de F\'{\i}sica, Universidade
Federal Fluminense; Av. Litor\^{a}nea s/n, Boa Viagem,
Niter\'{o}i 24210-340, RJ, Brazil; pmco@if.uff.br

\noindent
$^2$ Visiting from Institute for Theoretical Physics, Cologne
University, D-50923 K\"oln, Euroland;
stauffer@thp.uni-koeln.de

\bigskip

Abstract: We present different bit-string models of haploid
asexual populations in which individuals may exchange part of
their genome with other individuals (parasex) according to a
given probability. We study the advantages of this parasex
concerning population sizes, genetic fitness and
diversity.  We find that the exchange of genomes always
improves these features.

\section{Introduction}

Since a long time the question on why sex evolved has been
studied through different models. Some of them justify the
sexual reproduction from intrinsic genetic reasons, and others
from extrinsic or social reasons like child protection,
changing environment or parasites \cite{jorge}.

The Redfield model \cite{redfield} is an example of an elegant
model that requires little computer time.  It is not a
population dynamics model following the lifetime of each
individual, but only simulates their probabilities to survive
up to reproduction. The mortality increases exponentially with
the number of mutations in the individual. For the sexual
variant the number of mutations in the child is determined by
a binomial distribution such that on average the child has as
its own number of mutations half the number of the father,
plus half the number from the mother. At birth, new mutations
are added following a Poisson distribution, for both asexual
and sexual reproduction. Because of the lack of an explicit
genome, intermediate forms of reproduction as meiotic
parthenogenesis and hermaphroditism were not simulated.

A more realistic model, involving an explicit genome in the
form of bit-strings, was more recently used by \"Or\c cal et
al. \cite{orcal} to investigate intermediate reproductive
regimes. It makes use of a parameter
$\mu$, introduced before by Jan et al. \cite{jan}, defined
such that only individuals with $\mu$ and more mutations
exchange genome.  Healthy individuals without many mutations
reproduce asexually, that is, by cloning plus deleterious
random mutations.  

The models we present here are of this second type, that is,
the genomes of the individuals are represented by bit-strings,
and our purpose is to investigate an intermediate
strategy (between asexual-haploid and sexual-diploid reproduction) 
which is called parasex (S. Cebrat, private communication). 

\section{Penna-type models}

\subsection{General}

For biological ageing, the Penna model \cite{penna} presently
is the most widespread computer simulation method. The genome
of each individual is given by a string of 32 bits,
representing dangerous inherited diseases (detrimental
mutations) for the at most 32 intervals of life of this
individual. A 0-bit means health, a 1-bit on position $a$ of
the bit-string means a mutation affecting the health from that
age $a$ on. Three such diseases kill the individual at that
age $a$ at which the third disease becomes active. At each
time step, i.e. one iteration of the whole population, 
each living individual above a
minimum reproduction age of 8 gives birth to three children
with the same genome as the mother except for one mutation:
One bit position is randomly selected and its bit is set to
one independently of its previous value. Besides these deaths
from genetic reasons, individuals also die at each time step
with the Verhulst probability $N/N_{max}$ where $N$ is the
total population and $N_{max}$ a constant parameter, often
called the carrying capacity. For more results from the Penna
model we refer to \cite{book,stauffer}. In particular, this
model was used to compare various forms of sexual and asexual
reproduction for diploids \cite{anais,ijmpc}. We start with
the asexual Fortran program published in \cite{book}.

\begin{figure}[hbt]
\begin{center}
\includegraphics[angle=-90,scale=0.5]{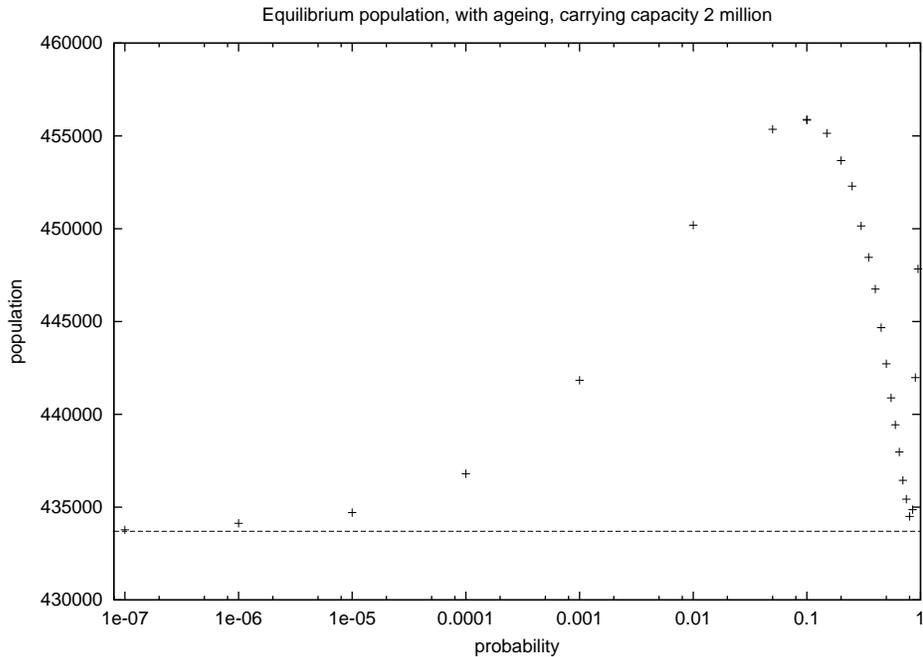}
\end{center}
\caption{
Variation of equilibrium population with parasex probability
$p$, including ageing. At least $10^4$ iterations at
$N_{max}= 2$ million; averages over second half of the
simulation. The horizontal line gives the result for the
standard Penna model without parasex and with otherwise the
same parameters.}
\end{figure}

\subsection{With ageing}

To simulate parasex, at each time step each individual gets
with probability $p$ from another randomly selected individual
half of its genome: For each of the 32 bits we determine
randomly if the old bit is kept or the bit of the other
individual is inserted instead. The number of deleterious
mutations (1-bits) is then found from the newly formed
bit-string.  This model might correspond to mutations of haploid
individuals hundreds of million years ago to evolve from asexual to
sexual reproduction. Bacteria to our present knowledge show no
ageing effects and are thus not described by this model but by
the one in the next subsection.

Fig.1 compares the population from the standard asexual Penna
model with that from this parasex model. This is a simple way
to compare the fitness of species under the same environment:
the larger the stationary population $N$ is for the same
$N_{max}$, the fitter is this species \cite{anais,ijmpc}. We
see a fitness maximum at a probability $p \simeq 0.1$. These
organisms also live longer with parasex than without (not
shown). With also positive mutations, however, the advantages of 
parasex are lost. A positive mutation means that when the randomly 
selected bit to be mutated is equal to one, it is set to zero in 
the offspring genome. 

\subsection{Without ageing}

Now we get rid of the ageing effects (increase of mortality
with increasing age) trying to model present-day bacteria. For
this purpose the program \cite{book} no longer counts at each
iteration the current number of active mutations; instead the
number of mutations is determined at birth (inherited
mutations plus new mutation happening at birth), and stays
with this individual until its death. Otherwise the program
remains essentially as published \cite{book}, i.e. again three
mutations kill.
By definition, the survival probabilities of all living individuals 
is independent of their number of mutations (0, 1 or 2), while the 
survival of their offspring depends on this number. 

\begin{figure}[hbt]
\begin{center}
\includegraphics[angle=-90,scale=0.5]{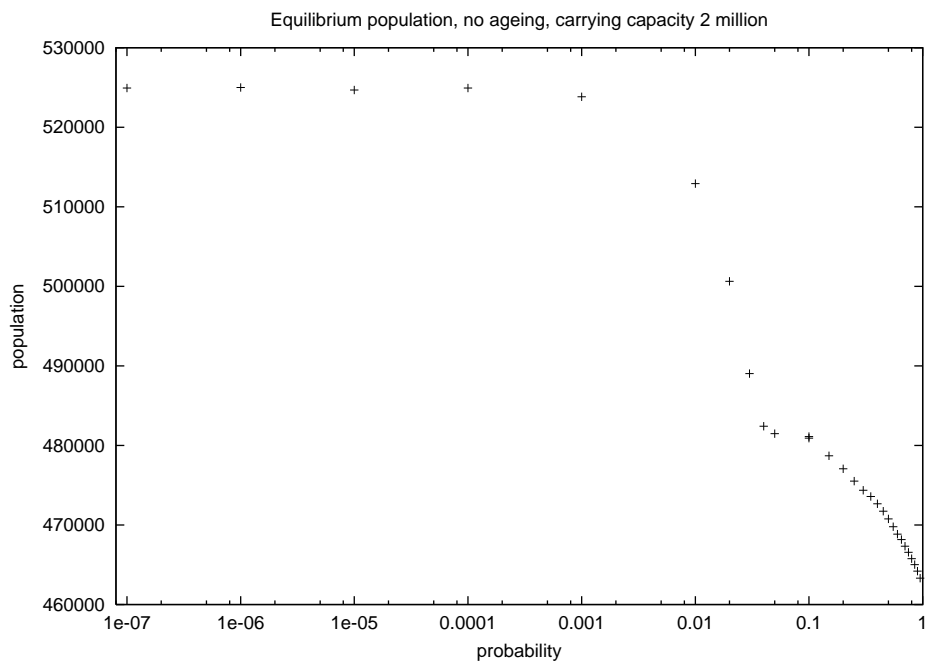}
\end{center}
\caption{
As Fig.1, but without ageing.
}
\end{figure}

``Death'' for bacteria is a matter of definition: If one
bacterium splits into two, we may regard this event as the
death of the previously living and the birth of two new
bacteria. Computationally less changes are required if we
define this duplication process as a mother giving birth to
one child. The mother then keeps the old genome, while the
child has at most one more mutation. In this way, the
individuals are genetically immortal and die only from
environmental causes simulated by the above mentioned Verhulst
factor. The mortality (probability to die within the time up 
to the next birth) is thus positive but independent of age. 

Fig.2 shows the stationary populations in this model without
ageing, but still with the above birthrate of three. (One
iteration may thus approximate two consecutive duplications.)
We see a fitness plateau at a low parasex probability up to
$10^{-3}$.

Simulations with birth rate = 1 instead of 3 are shown in
Fig.3, giving similar results. For both Fig.2 and 3 our model 
gives much lower populations (166032 and
45466, respectively), if we omit completely the exchange of genomes. 
We feel that this comparison, which makes parasex very advantageous, 
is the appropriate one. 

\begin{figure}[hbt]
\begin{center}
\includegraphics[angle=-90,scale=0.5]{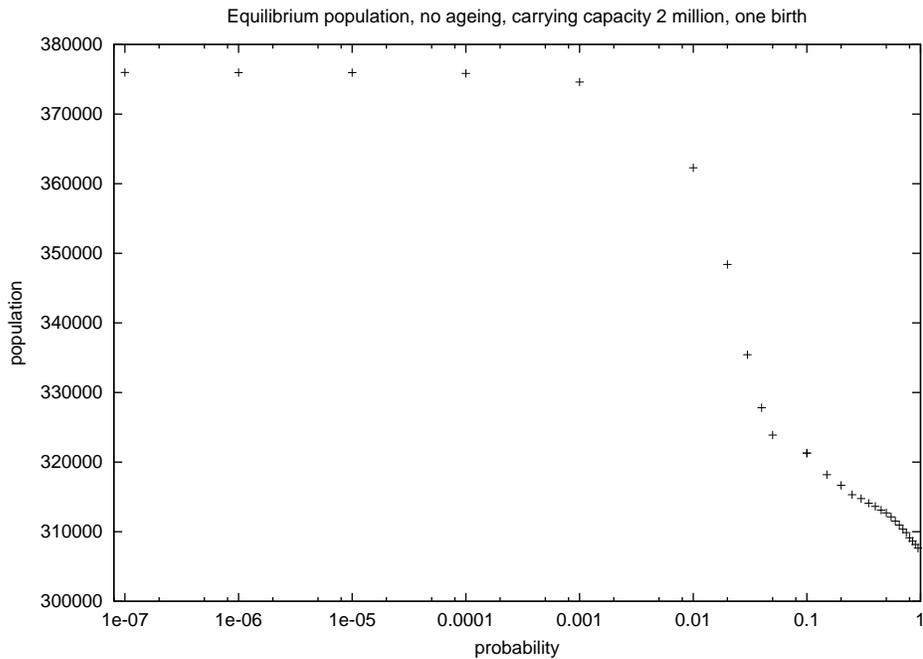}
\end{center}
\caption{As Fig.2, but with one instead of three births per
iteration.}
\end{figure}

\section{Bit-string model with longer genomes}  

\subsection{General} 

This bit-string model has no age structure, and was previously
introduced by de Oliveira \cite{pmco} for diploid populations
(see also \cite{mit},\cite{ticcona},\cite{js}). Here we
simplify it for haploids. The genome of each individual is
represented by a bit-string now of 1024 bits each. At every time 
step each
individual $i$ produces a copy of its own genome (child), and
mutations are randomly introduced to this copy according to a
predefined mutation rate $m$. Now mutations can be only
deleterious (the randomly selected bits are always set to one)
or {\bf mixed}. In the last case if a randomly selected bit of
the original genome is equal to one, it is set to zero in the
child genome (positive mutation) and vice-versa (bad mutation).  

The population size $P$ (the number of alive individuals) is
kept nearly constant, always fluctuating around some number
$P_0$, say $P_0 = 10000$, by applying the following death rule
after the whole population breeding is over. Each individual
$i$ survives with probability $x^{N_i + 1}$, where $N_i$
{\bf is the number of 1-bits} in its genome. Before killing
any individual, the value of $x$ is determined in order to
give an overall death rate that compensates the number of
newborns, that is, by solving the equation 
\begin{equation}
x \sum_{i=1}^{P} {x}^{N_i} = P_0 \,\,\,\, ,
\end{equation}     
where the sum runs over all individuals, including the
newborns. Once the value of $x$ is already known, {\bf then 
the death roulette is applied} sequentially to each individual $i$,
according to its $N_i$. In this way, the larger is the number of 
1-bits in a given individual's genome, the smaller is its survival 
probability. In contrast to the previous model (section 2.3), now 
individuals differ from each other only according to their own survival 
probabilities, the number of offspring per time step being the same 
for all.     

After the killing process described above, again for each
individual $i$, a random number between zero and one is
selected and compared to a predefined probability $p$ of
exchanging genomes. If this random number is smaller or equal
to $p$ a new individual $j$ is randomly selected from the same
population and the genomes of $i$ and $j$ are crossed. 
The same crossing position is randomly selected along the 
1024 bit-string for both, and one piece
of individual's $i$ genome is recombined with the
complementary piece of individual's $j$ genome. The same
recombination occurs with the remaining parts, producing in
this way two new genomes, without changing the total number of
individuals.  After completing the process of breeding,
killing and recombination on the whole population, one more
time step is added to the evolutionary clock of the whole
system.

\subsection{Results for mixed mutations} 

All results presented in this subsection were obtained for a population
size $P_0=10000$ individuals, a total number of 65 thousands
time-steps with averages performed during the last 32
thousands. At this stage, the probability distribution of
1-bits in the genomes of all individuals has already reached a
steady state. 
\bigskip

\bigskip

\begin{figure}[hbt]
\begin{center}
\includegraphics[angle=-90,scale=0.40]{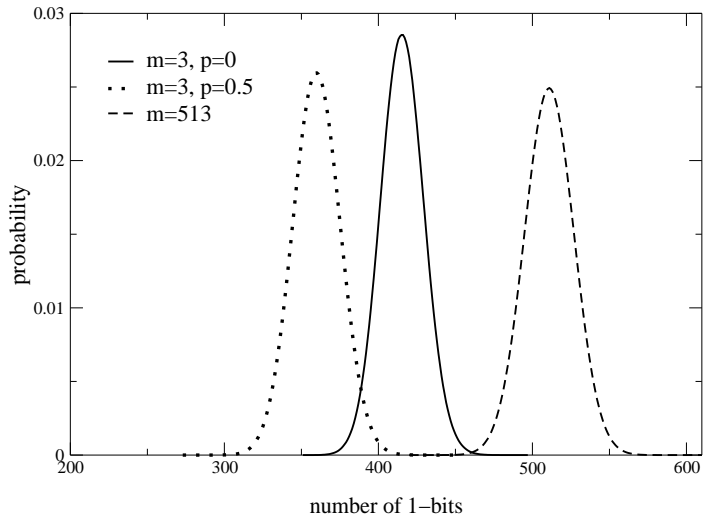}
\end{center}
\caption{ The probability distribution of 1-bits per individual, with 
($p=0.5$) and without parasex ($p=0$), for mutation rates $m=3$ and 
$m=513$ (in this last case, at right, the curves for $p=0$ and $p=0.5$ 
coincide).  
}
\end{figure}
  
First, we studied this distribution when there is no parasex.
We observed that for a very high mutation rate ($m \ge
512$), that is, if during the reproduction step half or more of 
the offspring genome is mutated, there is no selection effect:  
Genomes end
up, on average, with half of the bits equal to one (maximum
entropy). In this case the introduction of parasex has no
visible effect and the probability distribution of 1-bits per
individual seems to be a Gaussian centered at 512, in both
cases (rightmost curve of Fig.4.).
\bigskip

For small mutation rates, selection effects are already
present even without parasex: The probability distribution is
now centered at a value smaller than 512. The introduction of parasex 
moves the distribution towards an even smaller average number of
1-bits, as shown in Fig.4. 
\bigskip

In order to confirm that these distributions are Gaussians,
that is: $$y = A \exp \biggl [ {-\frac{(N - \langle N \rangle)^2}
{2 {\alpha}^2}} \biggr ] $$ where $\langle N \rangle$ is the average 
number of 1-bits and
$\alpha$ is the distribution width, we plot $Y = - \ln y$ as a
function of $X = (N - \langle N \rangle)^2$, to check if we obtain straight
lines for both sides of the distribution. The slope of these 
lines is proportional to $1/{\alpha^2}$. The results are
shown in Fig.5, for a mutation rate of $m=513$.
\bigskip

\begin{figure}[hbt]
\begin{center}
\includegraphics[angle=-90, scale=0.40]{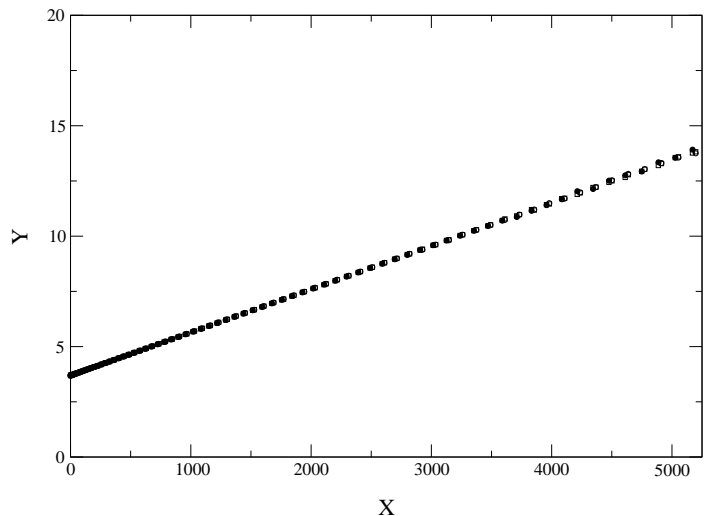}
\end{center}
\caption{ Linearization (see text) of the distributions with 
$m=513$ shown in Fig.4. 
}
\end{figure}

In this case where, as mentioned before, selection does not work, 
it can be seen that the distributions are indeed Gaussians. 
\bigskip

However, from Fig.6 we can see that for small mutation rates  
($m=3$) the tails of the distributions deviate from
the Gaussian form, mainly when there is no parasex (circles).
It occurs because selection is working and there are more
individuals populating the left side of the distributions, than
the right one. The reason why this deviation is stronger when
there is no crossing is that parasex tends to mix the very
clean genomes already selected by the dynamic process.
However, this effect should not be confounded with a possible
loss of diversity. On the contrary, besides improving the
average genetic fitness of the whole population by pushing the
distribution to the left, as shown in Fig.4, parasex in fact 
increases the
diversity, i.e. the width of the distribution. Note the lower 
slope of the lines in Fig.6 when there is parasex. The role of
the crossing mechanism is simply to offer a larger number of
options to the Darwinian selection process.
\bigskip

\begin{figure}[hbt]
\begin{center}
\includegraphics[angle=-90,scale=0.45]{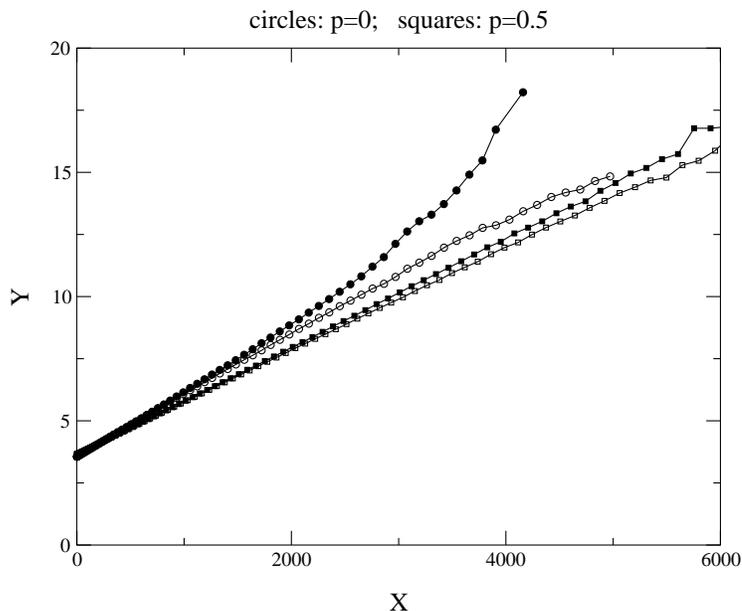}
\end{center}
\caption{ Linearization (see text) of the distributions with 
$m=3$, shown in Fig.4. 
The upper line corresponds to the right part of the distribution and the 
lower line to the left part. Circles: no parasex; Squares: 
parasex for $p=0.5$. 
}
\end{figure}  

Similar results were obtained for smaller values of the parasex 
probability $p$. Also, in all cases the final steady state distribution 
of 1-bits is reached much faster when parasex is allowed. This is another 
advantage in what concerns the ability of the population to adapt to 
sudden environment changes.

\subsection{Results for pure bad mutations}

When only deleterious mutations are considered, without parasex 
all genomes end up with only 1-bits, independently of the value of the 
mutation rate. For real genomes this would 
result in a genetic population meltdown with the death of 
all individuals. The introduction of parasex is able to avoid this 
extinction, even for small probabilities of exchanging genomes.  
Fig.7. shows the probability distribution for genome lengths
of 512 bits. Without parasex the distribution evolves to a delta-function 
at 512.    
\bigskip

\bigskip

\begin{figure}[hbt]
\begin{center}
\includegraphics[angle=-90,scale=0.38]{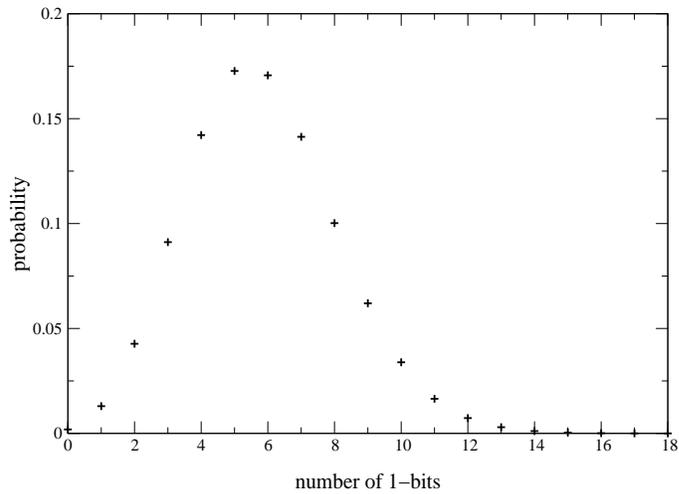}
\end{center}
\caption{ The probability distribution of 1-bits per individual, with parasex    
and only bad mutations. Parameters: $p=0.3$, $m=1$ and $P_0=1000$.
}
\end{figure} 

\section{Conclusions} 

We have used three different models to simulate genome exchange 
between individuals, a possible intermediate behaviour between 
asexual and sexual 
reproduction. In all of them the genome of each individual is  
represented by one bit-string (haploid) and its fitness is related to 
the number of 1-bits accumulated in its genome. Every individual has a 
probability to exchange part of its genome with another individual   
randomly chosen among the same population, a process called parasex. 
We have observed that this exchanging of genomes always increase the survival 
chances, measured through the population size or through the probability 
distribution of 1-bits per genome, depending on the model. 
Even in extreme cases where the population evolves to genetic meltdown, 
the introduction of parasex is enough to avoid it. 

\bigskip

\noindent {\bf Acknowledgements}: To PMMH at ESPCI for the warm hospitality,  
to Sorin T\u{a}nase-Nicola for helping us with the computer facilities and 
to S. Cebrat for discussions; SMO and PMCO thank the Brazilian agency FAPERJ 
for financial support.


\newpage

\begin{thebibliography} {99}

\bibitem{jorge} J.S. Sa Martins: Phys. Rev. E 61, 2212 (2000);
R.S. Howard and C.M. Lively, Nature 367, 554 and 368, 358 (E)
(1994).

\bibitem{redfield} R.J. Redfield: Nature 369, 145 (1994).

\bibitem{orcal} B. \"Or\c cal, E. T\"uzel, V. Sevim, N. Jan
and A. Erzan: Int. J. Mod. Phys. C 11, 973 (2000).

\bibitem{jan} N. Jan, L. Moseley and D. Stauffer: Theory in
Bioscience 119, 166 (2000).

\bibitem{penna} T.J.P. Penna: J. Stat. Phys. 78, 1629 (1995).

\bibitem{book} S. Moss de Oliveira, P.M.C. de Oliveira, D.
Stauffer: {\it Evolution, Money, War and Computers}, Teubner,
Stuttgart and Leipzig 1999.

\bibitem{stauffer} D. Stauffer, page 258 in: {\it Biological
Evolution and Statistical Physics}, ed. by M. L\"assig and A.
Valleriani, Springer, Berlin Heidelberg 2002.

\bibitem{anais} D. Stauffer, P.M.C. de Oliveira, S. Moss de
Oliveira, T.J.P. Penna, J.S. Sa Martins: An. Acad. Bras. Ci.
73, 15 (2001) (condmat/0011524).

\bibitem{ijmpc} D. Stauffer, J.S. Sa Martins, S. Moss de
Oliveira: Int. J.  Mod. Phys. C 11, 1305 (2000); J.S. Sa
Martins, D. Stauffer: Physica A, 294, 191 (2001).

\bibitem{pmco} P.M.C. de Oliveira: Theory Bioscience 120, 1
(2001) (condmat/0101170).

\bibitem{mit} P.M.C. de Oliveira, S. Moss de Oliveira and Jan
P. Radomski: Theory in Bioscience, 120, 77 (2001).

\bibitem{ticcona} A. Ticona and P.M.C. de Oliveira: Int. J.
Mod. Phys. C 12, 1075 (2001).  

\bibitem{js} Jan P. Radomski and S. Moss de Oliveira: Int. J.
Mod. Phys. C 11, 1297 (2000).

\end{thebibliography}
\end{document}